\documentstyle[preprint,aps]{revtex}

\tighten
\begin{document}
\draft
\title{Dealing with the center and boundary problems\\ 
	in 1D Numerical Relativity}
\author{A. Arbona, C. Bona}
\address{
Departament de F\'{\i}sica, Universitat de les Illes Balears.\\
E-07071 Palma de Mallorca. Spain}

\author{Corresponding author: Carles Bona}
\address{
Departament de F\'{\i}sica, Universitat de les Illes Balears.\\
E-07071 Palma de Mallorca. Spain\\
dfscbg0@ps.uib.es\\
telephone: +34-971 173222\\
fax number: +34-971 173426\\}

\maketitle

\begin{abstract}
Instabilities in finite difference codes due to the singularity of
spherical coordinates at the center are studied. In typical Numerical
Relativity applications, standard regularization techniques by
themselves do not ensure long term stability. A proposal to remedy
that problem is presented, which takes advantage of redundant
quantities introduced in recent hyperbolic formulations of Einstein's
evolution equations. The results are discussed through the example
case of a boson star, where a significant improvement in the
implementation of boundary conditions is also presented.
\end{abstract}


\narrowtext

\section{INTRODUCTION}

Spherically symmetric systems have deserved a lot of interest since the early
stages of Numerical Relativity. In the pioneering work of May and
White\cite{May}, the evolution of a self-gravitating spherical star was 
modelled by coupling the hydrodynamical equations with two first integrals of 
Einstein field equations: the energy and momentum constraints. These
are two ordinary differential equations, 
which in spherical symmetry are enough
to determine the evolution of the whole metric. The validity of this
approach, however, is restricted to the spherical
case, because in more general situations these first integrals are not enough
to determine the solution of Einstein field equations. Furthermore,
even in the spherical case, the treatment of boundary conditions with
this system is known to be
rather difficult. The latter is especially dangerous in systems with
outgoing matter or
radiation, which need to be carefully dealt with.

The alternative approach is to couple the hydrodynamical equations with
the full set of Einstein partial differential equations. In that
way one must evolve both the hydrodynamical
quantities and the spacetime metric on the same footing. The resulting 
formalism can then be extended in principle to axially symmetric (2D)
or even completely general situations. 
In the spherically symmetric case, however, the major drawback of this
approach is not just the higher number of metric 
quantities to compute. It is rather the fact that the spherical coordinate 
system is singular at the origin, even when the spacetime itself is regular.
The region close to the center of symmetry poses serious problems to numerical
codes, which are to be overcome by using special techniques like artificial
viscosity or algorithm matching. In most cases, this is usually done by 
trial and error and, together with the treatment of the outer boundary, becomes
the main limiting factor when constructing finite difference codes.

The purpose of this paper is to present a new way of writing
the full set of Einstein equations in spherical symmetry (with a
matter content),
which takes advantage of some
redundant quantities that appear in recent hyperbolic formulations of
Numerical Relativity~\cite{bonamasso92,bmss95}, in order to solve this
problem. In that way, one will be able to
treat the center region with exactly the same numerical algorithms as
any other region of the numerical grid, with a significative reduction
of the time devoted to code development.
The same formalism also allows us to impose more
accurate external boundary conditions, related to the physics
of the problem. The combination of these two improvements (center and
boundary treatment) helps in increasing the algorithm
stability and maintaining the required accuracy. As far as tested, the
resultant code does not show any stability problem, even after very
long runs.
We will work out the description of the method through the case of a
complex massive scalar field describing a
radiating boson star, where we easily reproduce previous results from other
authors\cite{seidelsuen}.

\section{FORMALISM}

The object of Numerical Relativity is to solve the Einstein equations
\begin{equation}
G_{\mu\nu}=8\pi G\:T_{\mu\nu},  \label{Eq.Einstein}
\end{equation}
in an approximate way. This is necessary when we deal with some of the
most interesting systems in astrophysics: black holes, neutron stars,
close binaries, supernovae, etc.\\ 
$G_{\mu\nu}$ is the Einstein tensor associated to a certain spacetime
(represented by a metric $g_{\mu\nu}$), $G$ is the gravity constant
and $T_{\mu\nu}$ is the momentum-energy tensor describing matter in
this spacetime.\\       
The metric
element can be written, in normal coordinates, as
\begin{equation}
 ds^2=g_{\mu\nu}dx^{\mu}dx^{\nu} = 
      -\alpha^2 dt^2+\gamma_{ij}\;dx^i\;dx^j    \label{metrica}
\end{equation}
where $\gamma_{ij} = g_{ij}$ is the space metric and $\alpha$ is the lapse, 
a gauge degree of freedom.\\
The Einstein system (\ref{Eq.Einstein}) is constituted by 10 coupled 
second order partial 
differential equations, when written in terms of the 4-dimensional metric 
$g_{\mu\nu}$. Six of them are evolution equations and the four remaining are 
constraint equations~\cite{lich,choquet,ADM}.\\
We deal with spherical symmetry problems, so the system reduces to a 
one-dimensional (1D) problem. In spherical coordinates, (\ref{metrica}) 
is written
\begin{equation}
    -\alpha^2 dt^2+g_{rr}dr^2+ g_{\theta\theta}\:d\Omega  \label{metrik}
\end{equation}
where $d\Omega = d\theta^2 + sin^2 \theta\; d\phi^2$.\\
The Einstein system can be written in first order form by introducing new
variables. For simplicity, we will use the vacuum equations to
illustrate our procedures. Following~\cite{bonamasso92}, we will
express it as it follows:
\begin{equation} 
   \partial_t g_{\theta\theta} 
	=  g_{\theta\theta} (C q^\theta{}_\theta )  \;,\\  \label{old_gzz}
\end{equation}
\begin{equation}                                                              
   \partial_t g_{rr} 
   		=  g_{rr} (C q^r{}_r )  \;,\\     \label{old_grr}
\end{equation}
\begin{equation}
   \partial_t C  = C\left(1/2\;C[(f-1)q^r{}_r+2f\;q^\theta{}_\theta]
	\right) \;,\\  \label{old_c}
\end{equation}
\begin{equation} 
   \partial_t L_r - \partial_r\left(
   C\;f [q^\theta{}_\theta + 1/2\;q^r{}_r]\right)  =  0 \;,\\   
	\label{old_lr}
\end{equation}
\begin{equation} 
   \partial_t D_{r\theta}{}^\theta
- \partial_r( C q^\theta{}_\theta)  =  0 \;,\\
	   \label{old_dzz}
\end{equation}
\begin{equation}
   \partial_t trD-\partial_r\left(C(q^r{}_r+2q^{\theta}{}_{\theta})
     \right) = 0\\  \label{old_trd}\;,\\
\end{equation}
\begin{equation} 
   \partial_t V_r    =  2C \left[q^\theta{}_\theta\;L_r
  - 1/2 (q^\theta{}_\theta-q^r{}_r)D_{r\theta}{}^\theta \right] \;,\\
	 \label{old_vr}	
\end{equation}
\begin{eqnarray}
   \partial_t (q^r{}_r)&-& \partial_r\left(
		2C[3/4\;V_r - D_{r\theta}{}^\theta + L_r]\right) =\nonumber\\
	   &-&C \biggl(1/2\;q^\theta{}_\theta(q^r{}_r-q^\theta{}_\theta) 
            -\frac{g_{rr}} {g_{\theta\theta}}\nonumber\\
	   &+& L_r\;D_{r\theta}{}^\theta
           +1/4[(q^\theta{}_\theta)^2-(D_{r\theta}{}^\theta)^2] \biggr) \;,
	   \label{old_qrr}
\end{eqnarray}
\begin{eqnarray}
   \partial_t (q^\theta{}_\theta)&-&
		\partial_r\left(
		C[D_{r\theta}{}^\theta - 1/2\;V_r]\right) =\nonumber\\ 
	&-& C \biggl(1/2\;q^\theta{}_\theta(q^\theta{}_\theta-q^r{}_r) 
         +\frac{g_{rr}} {g_{\theta\theta}} \nonumber\\
	&-& L_r\;D_{r\theta}{}^\theta
        +1/4[(q^\theta{}_\theta)^2-(D_{r\theta}{}^\theta)^2] \biggr) \;
     \label{old_qzz}
\end{eqnarray}
where we have defined 
$L_r \equiv \partial_r ln(\alpha)$,
$D_{rr}{}^r \equiv \partial_r ln(g_{rr})$,
$D_{r\theta}{}^\theta \equiv \partial_r ln(g_{\theta\theta})$,
$trD \equiv D_{rr}{}^r+2D_{r\theta}{}^\theta$,
$C\equiv{\alpha \over \sqrt{g_{rr}}}$ (speed of light),
$q^r{}_r \equiv {1 \over C} {\partial_t g_{rr} \over g_{rr}}$ and 
$q^{\theta}{}_{\theta} \equiv {1 \over C} {\partial_t g_{\theta\theta} 
\over g_{\theta\theta}}$. Also, allowing for the 
momentum constraint, $V_r=2D_{r\theta}{}^\theta$.\\
Note that we have written the Einstein system as a first order balance law
system. The variable $V_r$ is redundant: it was introduced with the
purpose of making the system hyperbolic, which allows us to directly
deal with diagonalized variables with a well-defined speed and sense
of propagation (very useful when imposing boundary conditions, as we
will see later).

The function $f(\alpha)$ 
determines the gauge.
We have chosen the gauge degree of freedom $\alpha$ to follow the harmonic 
slicing~\cite{bonamasso88}, which is the simplest choice ($f=1$) 
and good enough for singularity-free spacetimes.\\

\section{Geometrical factors and numerical error}

Due to spherical symmetry $g_{\theta\theta}$ has a strong geometrical factor:
it is proportional to $r^2$ when $r \rightarrow 0$. 
It follows that both $D_{r \theta}{}^{\theta}$ and $V_r$ are proportional 
to $1/r$ , a singular behaviour at the origin.\\
Simulations dealing directly with these equations, or equivalent ones,
are extremely unstable. This is well-known and
some alternatives are usually taken.
In previous works dealing with black holes, the problem was avoided by
putting internal boundary conditions at the apparent horizon. In this way
the grid was cut at the apparent horizon, far enough from both the 
physical and 
coordinative singularity at $r=0$. The problem is completely different
when we deal with singularity-free systems (stars). There, there is no
way of cutting the solution before $r=0$ without loosing physically
relevant information.
The coordinative singularity at $r=0$ turns up as a big danger for 
accuracy and stability near the origin when using the standard system 
(\ref{old_gzz}-\ref{old_qzz}).\\
The first step for dealing with this is to analytically extract the
geometrical factors, as far as possible, from the equations, so they
only deal with the regular part. This is done by writing the line 
element as
\begin{equation}
  -\alpha^2 dt^2+g_{rr}\:dr^2+r^2 {\tilde g_{\theta\theta}}\:d\Omega  
	\label{new_metrik}
\end{equation}
so that the system, in our particular case, is rewritten with 
$g_{\theta\theta} = r^2 {\tilde g_{\theta\theta}}$,
$D_{r \theta}{}^{\theta} = {1\over r} + {\tilde D_{r \theta}{}^{\theta}}$,
${tr\tilde D} = D_{rr}{}^r+2{\tilde D_{r\theta}{}^\theta}$. The new
equations read
\begin{equation}
\partial_t {\tilde g_{\theta\theta}}={\tilde g_{\theta\theta}}\:C\:
	q^{\theta}{}_{\theta}  \\   \label{mod_gzz}
\end{equation}
\begin{equation}
\partial_t {\tilde D_{r\theta}{}^{\theta}}-
\partial_r\left(C q^{\theta}{}_{\theta}
     \right)=0 \\   \label{mod_dzz}
\end{equation}
\begin{equation}
   \partial_t {tr\tilde D}-\partial_r\left(C(q^r{}_r+2q^{\theta}{}_{\theta})
     \right) = 0\\  \label{mod_trD}
\end{equation}

This is yet an improvement because the evolved quantities
${\tilde D_{r\theta}{}^{\theta}}$ and ${\tilde g_{\theta\theta}}$
are now regular at the center.
But further improvement should be achieved, provided we want to deal with a
stable code,
because rewriting (\ref{old_gzz}-\ref{old_qzz}) with (\ref{new_metrik}) 
introduces explicit $1/r$ factors in the terms
containing space derivatives (flux terms) and $1/r^2$ factors 
show up in the right hand side (source terms) of equations (\ref{old_qrr})
and (\ref{old_qzz}). 
We have somehow explicitly segregated the 
geometrical term, but it is still not out of the system. Now we can
appreciate the kind of error that is introduced: both fluxes and sources
are singular due to $1/r$ and $1/r^2$ terms and exact 
cross-cancellation is needed between them to keep the system regular.
However, this is not achieved by the finite difference algorithm due to the
truncation error when calculating the space derivatives of the fluxes.
The truncation error when computing a given variable with a second order
code is proportional to $(\Delta r)^2$, where $\Delta r$ is the distance
between neighbouring points in the grid. The terms $1/r$ in the fluxes give 
errors like $(\Delta r)^2/r^4$.
Where $r \rightarrow 0$ ($r \sim \Delta r$) the code does not converge.\\
As a particular instance we tried to solve the boson star 1D problem
\cite{seidelsuen} by
using these modified quantities. The numerical truncation error 
caused an instability that
prevented us from obtaining any meaningful result before the code
crash.\\
These are well-known regularization techniques to handle this
problem. In the next section, however, we
present a new approach which regularizes the equations in a deeper
way.\\

\section{The role of the momentum constraint}

Our proposal goes beyond the standard regularization procedures and
succeeds in removing the most
dangerous remaining $1/r$ factors. We take advantage
from the way in which the momentum constraint was used to build the
system (\ref{old_gzz}-\ref{old_qzz}). Remember that the variable $V_r$ is 
redundant and it was introduced with
the objective of making the system hyperbolic~\cite{bonamasso92}. 
As said, in (\ref{old_gzz}-\ref{old_qzz}) the 
momentum constraint is satisfied if and only if  
$V_r = 2 D_{r \theta}{}^{\theta}$. 
We can monitorize the level of violation of the momentum constraint
by testing this algebraic identity through the
evolution.\\
From the considerations of the preceding section, it will be natural
to evolve instead of $V_r$ the regular quantity 
${\tilde V_r}\equiv 2{\tilde D_{r\theta}{}^{\theta}}$. But it turns out to be 
more convenient to define instead
\begin{equation}
 {\tilde V_r}\equiv 2{\tilde D_{r\theta}{}^{\theta}}+{4\over r}
 \left(1-\sqrt{g_{rr}\over
 {\tilde g_{\theta\theta}}}\right)   \label{vrtilde}
\end{equation}
which is also regular (provided that 
$g_{rr}/{\tilde g_{\theta\theta}} - 1 \rightarrow 0$ as $r^2$, which
demands smooth initial data near the 
origin) and removes both the $1/r$ terms in the fluxes and the $1/r^2$ 
terms in the sources. The singularities
have now disappeared from all the evolved variables as well as the numerical
error caused by the incomplete cross-cancellation between fluxes and 
sources.\\
The modified equations are
\begin{equation}
\partial_t {\tilde V_r} = C\left(2 q^{\theta}{}_{\theta} L_r+(q^r{}_r-
      q^{\theta}{}_{\theta}){{\tilde V_r}\over 2}
     \right)\\      \label{mod_vr}
\end{equation}
\begin{eqnarray}
\partial_t q^r{}_r&-&\partial_r\left(2C({3\over4} {\tilde V_r}-{\tilde 
D_{r\theta}{}^{\theta}}+
     L_r)\right)=\nonumber\\
	C&\biggl(&{ q^{\theta}{}_{\theta}\over 2}
     \left({1\over 2}q^{\theta}{}_{\theta}-q^r{}_r\right)+{1\over 16}{\tilde 
V_r}{}^2+
	\nonumber\\
    && {\tilde D_{r\theta}{}^{\theta}}({{\tilde V_r}\over 2}-
	{\tilde D_{r\theta}{}^{\theta}})+
     2L_r {\tilde D_{r\theta}{}^{\theta}}\nonumber\\
    && -{3\over 2}L_r {\tilde V_r}+ 
     {2\over r}(4{\tilde D_{r\theta}{}^{\theta}}-{{\tilde V_r}\over 2}-{tr\tilde 
D})
	\biggr)   \label{mod_qrr}
\end{eqnarray}
\begin{eqnarray}
\partial_t q^{\theta}{}_{\theta}&-&\partial_r\left(C\left(
     {\tilde D_{r\theta}{}^{\theta}}-{{\tilde V_r}\over 2}\right)\right)=
	\nonumber\\
	C&\biggl(&-{
     q^{\theta}{}_{\theta}\over 2}\left({3\over 2}q^{\theta}{}_{\theta}-
     q^r{}_r\right)+L_r{{\tilde V_r}\over 2}+  
    { \left({\tilde D_{r\theta}{}^{\theta}}\right)^2\over 2}-\nonumber\\
    && {1\over 16}{\tilde V_r}{}^2+
     {1\over r}(2L_r+4{\tilde D_{r\theta}{}^{\theta}}-{tr\tilde D})
     \biggr) \label{mod_qzz}
\end{eqnarray}
There is still a point to be taken into account. In eqs. 
(\ref{mod_qrr}-\ref{mod_qzz}) the sources contain terms like
$1/r$ times ${tr\tilde D}$, $L_r$ or other variables that are radial
derivatives of the metric coefficients. 
These terms do not pose any problem as $r \rightarrow 0$, as radial 
derivatives of any differentiable function vanish at the origin.
However, in the finite difference approach, we cannot put
a grid point at $r=0$ due to the explicit $1/r$ factors. We will
therefore place the first grid point at $r=\Delta r/2$. 
With this new way for writing the Einstein equations in spherical
symmetry we are able to evolve the whole grid, without needing to
apply special techniques or different algorithms to the center region.
The equations are now intrinsically stable and can be directly coded.

\section{Example: the Boson Star}

We have applied this idea (\ref{vrtilde}) to a Boson Star
model~\cite{seidelsuen}.
Mathematically, the system is constituted by the Einstein equations
plus the Klein-Gordon ones. It has stable configurations known as
Boson Stars, which determine a mass-radius curve, as in the neutron
star case, with a local maximum for the star masses.\\
The stars located at the left (S-Branch) from this maximum (small radius) are
known to be stable against small perturbations. The stars at the 
right (U-Branch) are, on the other hand, unstable.
An U-Branch star, when perturbed can migrate towards a stable 
configuration in the S-Branch. The migration produces strong periodical
oscillations in the star radius around the final position, as well as
scalar field radiation.\\    
We tested our equations against this phenomenon. We
could not evolve the system with (\ref{old_gzz}-\ref{old_qzz}) far 
enough to measure any
relevant physical quantity. With (\ref{old_grr}-\ref{old_lr},
\ref{mod_gzz}-\ref{mod_trD}, \ref{mod_vr}-\ref{mod_qzz}),
we have been able to 
reproduce the migration and the complex behaviour of its stationary
states, with run time just limited by the required accuracy and not by
algorithm stability, which is not easily achieved with the alternative
formulation. We have to point out that part of the success comes from the
hyperbolic balance law formulation itself, by allowing more physically
consistent external boundary conditions. In the next section
we go further in that point.
In figure 1 we show the migration from an unstable configuration to 
a stable one~\cite{seidelsuen}. With the original form of 
(\ref{old_gzz}-\ref{old_qzz}) we were unable to see nor 
half an oscillation.

\section{The boundary conditions}
We illustrate the method for boundary conditions by describing the
treatment of the Klein-Gordon equations in our example system.\\
Let's take a look at the equation for the real part, $\phi_1$, of the
Klein-Gordon field ($g^{\mu\nu}\phi_{;\mu\nu}=m^2 \phi$, in tensor
form):
\begin{eqnarray}
\partial_t \rho_1 - \partial_r (C\psi_1) &=& 0\\
\partial_t \psi_1 - \partial_r (C\rho_1) &=& C \left[
\left(D_{r\theta}{}^{\theta} + {2\over r} \right)\rho_1 -
q^{\theta}{}_{\theta} \psi_1 - m^2 g_{rr} \phi_1 \right]
\end{eqnarray}
where $\rho \equiv \partial_r \phi$, $\psi\equiv{1\over C}\partial_t\phi$.
These equations, like the rest of the system, are written in balance
law form. The flux terms are those that would appear in a flat space
wave equation (transport terms). The $m^2$ term in the right hand side
is an oscillatory term which comes from the $m^2 \phi$ in the Klein-Gordon
equation. The rest of the source terms are due to spherical
coordinates and the coupling between the Klein-Gordon field and the
spacetime metric.\\
To put a condition over the transport terms would be quite easy, but
the combination with the source terms makes the problem more
difficult. What one
would like to do is to treat separately each term of the
equation. Fortunately, this is possible in a numerical code, by using 
the 'operator splitting' approach. With the
Strang splitting (second order operator splitting) each step in the
numerical evolution can be decomposed in the following way:
\begin{equation}
\epsilon(\Delta t) = {\cal S}({\Delta t\over 2}) {\cal R}(\Delta t)
{\cal S}({\Delta t\over2})
\end{equation}
The whole step, $\epsilon(\Delta t)$, is split into three sequential
processes. The first one is ${\cal S}({\Delta t\over 2})$. There we
evolve during half a step only the 'source' terms in the equations. If
we restrict to the Klein-Gordon part, we have:
\begin{eqnarray}
\partial_t \rho_1 &=& 0\\
\partial_t \psi_1 &=& C \left[
\left(D_{r\theta}{}^{\theta} + {2\over r} \right)\rho_1 -
q^{\theta}{}_{\theta} \psi_1 - m^2 g_{rr} \phi_1 \right]
\end{eqnarray}
This is an ordinary differential equation system for each point in the
net. In this part there is no connection between points. Each point
evolves by itself. So we do not have to state any boundary condition
in this substep.\\
In the second substep, ${\cal R}(\Delta t)$, we only evolve the
transport part (during a whole time step), so we have
\begin{eqnarray}
\partial_t \rho_1 - \partial_r (C\psi_1) &=& 0\\
\partial_t \psi_1 - \partial_r (C\rho_1) &=& 0
\end{eqnarray}
The remaining terms here are the wave equation-like ones. We know that
the solution is formed by two components ($\rho+\phi$ and
$\rho-\phi$), which propagate in opposite senses at speed $C$ (in
general we need the matter part to be written in hyperbolic form, so it
can be diagonalized).\\
Now here is where we need to put a boundary condition. What we want
is: first, the outgoing component not to be touched; any condition
over it would be superfluous or contradictory. Simply let it go
out. Second, as we suppose that there is no scalar field outside the net,
nothing is coming in (we disregard nonlinear effects between outgoing
radiation and external metric), so the boundary condition to impose is
simply
\begin{equation}
\rho_1+\psi_1=\rho_2+\psi_2=0  \label{bound}
\end{equation}
The third step is again ${\cal S}({\Delta t\over 2})$.\\
This is all. The only boundary condition we need to put is
(\ref{bound}), which is very simple indeed. Remember that there is not
gravitational radiation in spherical symmetry. The momentum constraint
plus the coordinate gauge condition fully determine the remaining
(metrical) part of the boundary conditions. Note that the relation of
$D_{r\theta}{}^{\theta}$ with $V_r$ holds when the momentum constraint
is satisfied, so we can use the computed values of $V_r$ at the
boundary to obtain the value of the incoming non-gauge field there.

\section{Conclusions}
The stability and accuracy near the origin of spherically symmetric
dynamical systems in Numerical Relativity is strongly improved by a new
formalism which provides an automatic care-free treatment
of the geometrical factors arising from spherical symmetry. These treatment
involves making use of the redundant variable $V_r$, previously introduced 
for other purposes (to ensure the hyperbolicity of the
system). The same formalism also allows boundary treatment based on
physical considerations and helps in increasing accuracy and
stability. That is, the same characteristic of our system, i.e. the
introduction of the new quantity $V_r$, allows us to simultaneously improve
the treatment of the two main problems in spherically
symmetric star-like systems: the center (through a redefinition of
$V_r$, which brings the system to a regular form) and the boundary
(through hyperbolicity achieved thanks to $V_r$, which lets us
distinguish between incoming and outcoming fields and treating them
according to physical criteria).

We have started from equations 
(\ref{old_gzz}-\ref{old_qzz}), which are just the 1D version of a 
general 3D system~\cite{bonamasso92,bmss95} which is actually being applied 
in multidimensional Numerical Relativity. Our point is that this
system, when rearranged to deal with the geometrical factors, works 
fine in the 1D case.\\
We are presently considering the axially symmetric case (2D), where 
again the 'redundant' quantities $V_i$ could be helpful to get rid of
the geometrical factors that arise from the coordinate singularity at the 
symmetry axis. But this is still an open question.\\

\acknowledgments
 This work is supported by the Direcci\'on General para 
Investigaci\'on Cient\'{\i}fica y T\'ecnica of Spain under project
PB94-1177 and the Conselleria d'Educaci\'o, Cultura i Esports of the
Govern Balear under grant of kind A7.

\begin{figure}
\caption{The central value of the scalar field is graphed
against time for an U-Branch star which has suffered a 
small perturbation (1\% reduction on its central field). The star migrates 
from its former U-Branch position (central field value equal to 0.4)
towards a S-Branch configuration 
corresponding to a central field value equal to 0.1, approximately. 
The migration consists in wide, diminishing oscillations of both the
central field and the star radius around their final equilibrium 
configuration.} 
\label{fig1}
\end{figure}

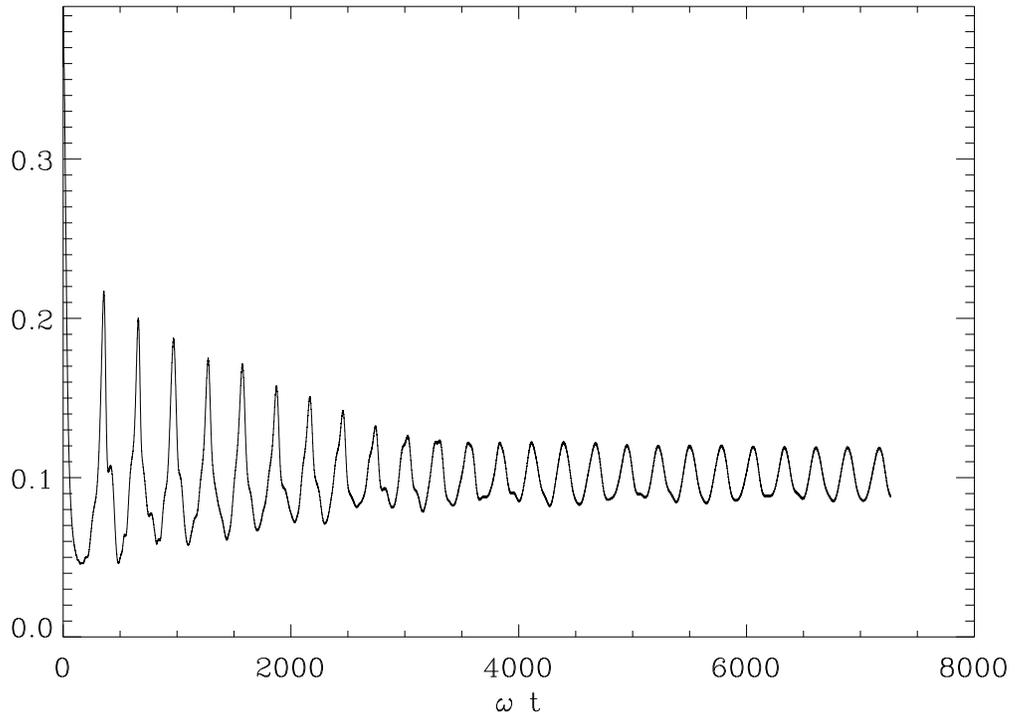
\end{document}